# Electrochemical Screening of Contact Layers for Metal Halide Perovskites

*Moses Kodur[1], Zachary Dorfman[1], Ross A. Kerner[2], Justin H. Skaggs[1], Taewoo Kim[1], Sean P. Dunfield[2-4], Axel Palmstrom[2], Joseph J. Berry[2,4,5], David P. Fenning[1]\**

[1]Chemical Engineering Program, Department of Nanoengineering, University of California, San Diego, La Jolla, California 92093, United States

[2]National Renewable Energy Laboratory, Golden, CO 80401, United States

[3]Materials Science and Engineering Program, University of Colorado, Boulder, CO 80309, USA

[4]Renewable and Sustainable Energy Institute, University of Colorado, Boulder, CO 80309, USA

[5]Department of Physics, University of Colorado, Boulder, CO 80309, USA

*Email: dfenning@eng.ucsd.edu

**Abstract**

Optimizing selective contact layers in photovoltaics is necessary to yield high performing stable devices. However, this has been difficult for perovskites due to their complex interfacial defects that affect carrier concentrations in the active layer as well as charge transfer and recombination at the interface. Using vacuum thermally-evaporated tin oxide as a case study, we highlight electrochemical tests that are simple yet screen device-relevant contact layer properties, making them useful for process development and quality control. Specifically, we show that cyclic voltammetry and potentiostatic chronoamperometry correlate to key performance parameters in completed devices as well as other material/interfacial properties relevant to devices such as shunt pathways and chemical composition. Having fast, reliable, scalable, and actionable probes of electronic properties is increasingly important as halide perovskite photovoltaics approach their theoretical limits and scale to large-area devices.

**TOC Graphic**

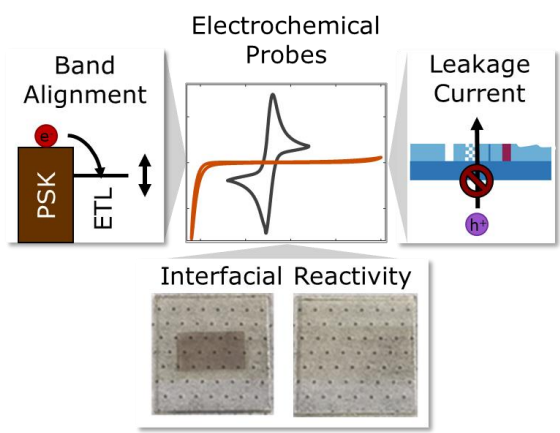

**Main Text**

High performance metal halide perovskite (MHP) photovoltaic devices ubiquitously employ heterostructure architectures where photogenerated charges are separated by carrier selective contacts.[1–3] A primary strategy to improve device efficiency is rooted in the development of carrier selective transport layers to control carrier extraction and process compatibility for a given perovskite composition.[4–13] For example, the electron selective transport layer (ESL) is chosen to minimize non-radiative recombination rates at the interface, align its conduction band minimum (CBM) with that of the absorber, and create a dense layer that minimizes leakage currents.[14,15] These properties are typically investigated by a suite of techniques including photoemission spectroscopy (PES), which is especially useful to determine the valence band maximum (VBM), surface chemistry, and fundamental relationships between the two.[16,17] Electrochemical techniques can provide similar, albeit indirect, information about these properties and have the advantage of being inexpensive, accessible, and compatible with large area or high throughput studies.[18–20] These characteristics make electrochemical measurements effective for transport layer development, monitoring process reproducibility, and implementing quality control, particularly at an industrial scale where the ultra-high vacuum requirements of PES complicate its implementation. Electrochemical characterization tools have been implemented to study other photovoltaic materials and devices, such as organic photovoltaics[21–24] and dye-sensitized solar cells,[25–28] but application to perovskites, and more specifically to perovskite solar cells, has been limited.[4,7,29] Here, we correlate electrochemical characteristics of a $SnO_x$ transport layer directly with perovskite photovoltaic performance parameters, demonstrating the utility of electrochemical probes for designing and characterizing perovskite photovoltaics.

In this work, we highlight the utility of electrochemical probes to rationalize the optimization of metal oxides for ESLs in optoelectronic devices.[30] Specifically, we use this approach to understand interface chemistry and energy level alignment dictated by processing dependencies of a vacuum thermally-evaporated $SnO_x$ (VTE-$SnO_x$) ESL used as the superstrate contact for MHP solar cells. We find that the cyclic voltammetry (CV) characteristics of the VTE-$SnO_x$ layers in an aqueous ferri-/ferrocyanide ($Fe[CN_6]^{3-/4-}$) electrolyte correlate to the final performance of the devices. In particular, the cathodic onset potential at which the current for the complexed $Fe^{3+} \rightarrow Fe^{2+}$ reduction "turns on" is related to the processing-dependent electron transfer of the VTE-$SnO_x$ and correlates to the solar cell open-circuit voltage ($V_{OC}$). Additionally, anodic sweeps directly probe electronic current leakage pathways irrespective of their origin (defects, pinholes, conductive filaments, impurities, *etc*.), and the magnitude of this loss mechanism is reflected in the short-circuit current density ($J_{SC}$). Finally, to show how these properties are affected by the VTE-$SnO_x$ chemistry, we present potentiostatic chronoamperometry in an aqueous methylammonium iodide electrolyte where proton-assisted reduction of $Sn^{II}$ defects to $Sn^0$ occur at the surface of the VTE-$SnO_x$ at -1.0 V versus Ag/AgCl. For optimally-annealed VTE-$SnO_x$ with more complete conversion to $Sn^{IV}$, reduction to $Sn^0$ does not onset until more cathodic potentials (< -1.3 V versus Ag/AgCl), in agreement with O:Sn ratios determined by X-ray photoemission spectroscopy (XPS). Thus, the electrochemical techniques provide key empirical information to optimize transport layer processing.

Importantly, these insights are provided prior to device completion and adaptable to high-throughput experimentation or large areas, making them viable for quality control, which we demonstrate by analyzing a 100 cm$^2$ VTE-$SnO_x$ ESL Many processing approaches compatible with the electrochemical methods described here have been demonstrated for $SnO_x$ in the literature, including spin-coat,[31,32] chemical bath deposition,[33,34] atomic layer deposition,[35–37] thermal evaporation,[38,39] and sputtering.[40,41] Using the electrochemical measurements described in this manuscript provides a rapid and straightforward means to screen newly-developed transport materials and assists in their co-optimization with active layer processing to ensure high yield and high performance critical for the commercialization of MHP-based photovoltaics and related optoelectronic devices.

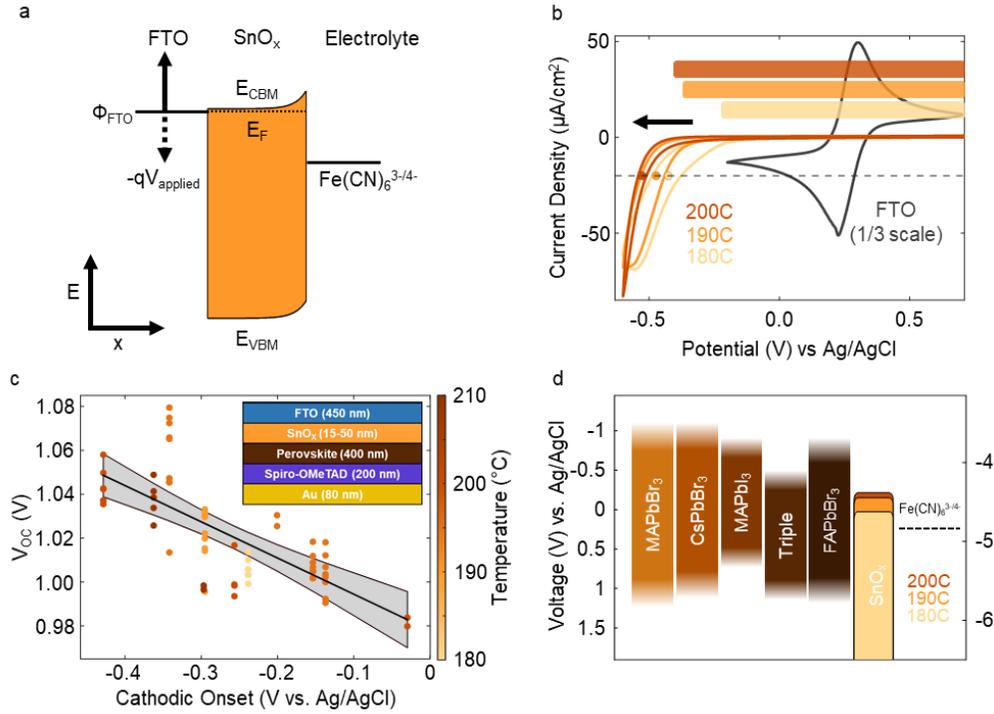

Figure 1. a) Schematic energy band diagram of the semiconductor-electrolyte junction formed between *n*-type VTE-$SnO_x$ and a ferri-/ferrocyanide redox couple in solution. b) Cyclic voltammograms (CV) of VTE-$SnO_x$ films annealed between 180 and 200°C show the blocking nature of the wide-gap VTE-$SnO_x$ until a large enough negative potential enables electron transfer from the VTE-$SnO_x$. For comparison, a CV of a bare FTO substrate, at one third scale, is shown (gray). The flat-band potential of each film is shown above the CV. c) The $V_{OC}$ of devices with varying VTE-$SnO_x$ is correlated with the cathodic onset potential extracted from CV, with a 95% confidence interval shown (device architecture in inset). The color bar indicates the post-deposition anneal temperature. For devices, the perovskite has a nominal stoichiometry of $Cs_{0.05}FA_{0.79}MA_{0.16}Pb(I_{0.84}Br_{0.16})_3$. d) A schematic alignment of the band edge positions of common perovskite chemistries compared with the flat-band voltages of the VTE-$SnO_x$ temperature series.[42,43] The dotted line indicates the standard redox potential of ferri-/ferrocyanide. From literature, this triple cation perovskite has a nominal stoichiometry of $Cs_{0.08}FA_{0.78}MA_{0.14}Pb(I_{0.86}Br_{0.14})_3$.

First, we demonstrate that CV in an electrolyte containing a reversible redox couple can be used to screen selective contacts (here VTE-$SnO_x$) for their suitability in MHP devices. In these measurements, thin-films of VTE-$SnO_x$ deposited on fluorine-doped tin oxide (FTO) substrates are placed in a 1.0 mM solution of the ferri-/ferrocyanide outer-sphere redox couple with a 0.5 M KCl supporting electrolyte, resulting in the formation of a semiconductor-electrolyte junction, as shown in Figure 1a (also see Supplementary Figure S1 and Note I). Without the VTE-$SnO_x$, the cyclic voltammogram on the bare metallic FTO electrode exhibits the classic duck-shaped curve of a reversible redox couple (Figure 1b).[44] In contrast, the band-bending induced in the VTE-$SnO_x$ (Figure 1a) leads to a sharp cathodic onset (Figure 1b). The cathodic onset shifts to more cathodic potentials with an increase in post-deposition annealing temperature from 180 to 200°C. These trends are representative of changes observed over a wider set of process variations shown in Figure 1c. Of process variations, annealing temperature had the strongest impact on the cathodic onset and thus $V_{OC}$, with a positive correlation between the two.

The onset potential of cathodic current is directly related to the VTE-SnO$_x$ chemistry and associated energetic alignment, which will impact the formation of the perovskite/ESL interface and ultimately correlates to the $V_{OC}$ of completed devices. Figure 1c shows the relationship between the cathodic onset voltage (defined as the potential at which current density = 20 µA/cm$^2$) and $V_{OC}$ of a completed FTO/VTE-SnO$_x$/FA$_{0.79}$MA$_{0.16}$Cs$_{0.05}$Pb(I$_{0.84}$Br$_{0.16}$)$_3$/2,2′,7,7′-tetrakis[N,N-di-p-methoxyphenylamino]-9,9′-spirobifluorene (spiro-OMeTAD)/Au device stack for several experiments with systematic variations made to the VTE-SnO$_x$ ESLs including thickness (10-50 nm), anneal temperature (180-210°C), and deposition rate (0.05-0.20 Å/s). The variability seen in the data set of Figure 1c illustrates that the reproducibility of the VTE-SnO$_x$/perovskite interface may be affected by multiple uncontrolled parameters in the processing of the VTE-SnO$_x$, such as ambient humidity and temperature.[45–48] (data shown in Table S1) A simple linear model fit to the data indicates with $p$-value $< 10^{-4}$ that a non-zero slope exists relating $V_{OC}$ to the cathodic onset potential. Despite this scatter, for the procedure used here, we observe that the post-deposition anneal temperature is the most sensitive tested processing parameter to manipulate the cathodic onset potential. Moreover, we find that regardless of the method used to define the onset potential,[49] the trend between $V_{OC}$ and the cathodic onset is consistent (Figure S2 and S3). Additionally, by comparing the results of CV to the flat-band potentials indicated by the horizontal bars in Figure 1b (determined by capacitance-voltage analysis of electrochemical impedance spectra, Supplementary Figure S4), we confirm that changes in the cathodic onset potential trend with the flat-band potential of VTE-SnO$_x$. Moreover, capacitance-voltage measurements indicate carrier concentration decreases from $10^{19}$ to $2\times10^{18}$ cm$^{-3}$ with increasing annealing temperature from 180 to 200°C (Figure S5). All films thus remain degenerately *n*-type. The resulting flat band potential is nearly equal to the CBM (Supplementary Figure S6 and Note II). Thus, the changes in the flat-band position correspond to a change in the CBM of the VTE-SnO$_x$ that occurs as the film is annealed in air. The qualitative energetic alignment is shown in Figure 1d alongside energy levels reported for various perovskite compositions including the "Triple" cation composition used in our devices.[42,43] Therefore, CV measures the combined effect of multiple electronic properties that dictate charge transfer at the interface, serving as a time-efficient early diagnostic for an improved interface (*e.g.* band alignment) between the ESL and perovskite absorbers.

We further leverage insight from the *anodic* portion of the CV, which serves as a measure of undesirable leakage currents, agnostic to their origin, and relates to $J_{SC}$. For example, in Figure 2a, cyclic voltammograms show that VTE-SnO$_x$ films of increasing thickness exhibit reduced anodic current. In perfect hole-blocking films there would be no anodic hole current because of the lack of states available for charge transfer. However, whether by charge transfer with the FTO at pinholes, via conductive filaments, or other defects (inset schematic Figure 2a), small hole currents can pass through the film. We further analyzed and attempted to quantify the pinhole density using chronoamperometry. (Figure S7) These small current pathways are also evidenced by anodic leakage current in CV and reduce the selectivity of the contact, thereby negatively affecting current collection and the $J_{SC}$ of completed devices (Figure 2b). These data also a possess a $p$-value $< 10^{-4}$ demonstrating a statistical correlation between J$_{SC}$ and the anodic current.). Additionally, we find that the shunt resistance is negatively correlated with the anodic current. (Figure S8) Note the 50 nm thick VTE-SnO$_x$ films in Figure 2b display variation in anodic leakage current magnitude and do not necessarily provide the lowest leakage currents, evidencing that processing parameters other than thickness must be considered to minimize anodic currents. ETL thickness modulated the leakage current the most of test process parameters, with intermediate thicknesses of ~20-30 nm producing the highest $J_{sc}$.

If leakage currents are assumed to be dominated by pinholes, the effective surface coverage of the ETL can be estimated by modeling the time-dependent change in current seen after a potential step based on diffusion-limited electron transfer at pinholes (Figure S7a).[44] In all cases there was a >97% reduction in the anodic current relative to when no ETL was present. Thick ETL films (at 34 nm) blocked 99.4% of the current (Figure S7b). Lower anodic currents thus appear unsurprisingly related to improved surface coverage, as has been shown previously in studies of titania for dye-sensitized solar cells,[5] and are indicative of lower shunt resistance in final devices (Figure S8).

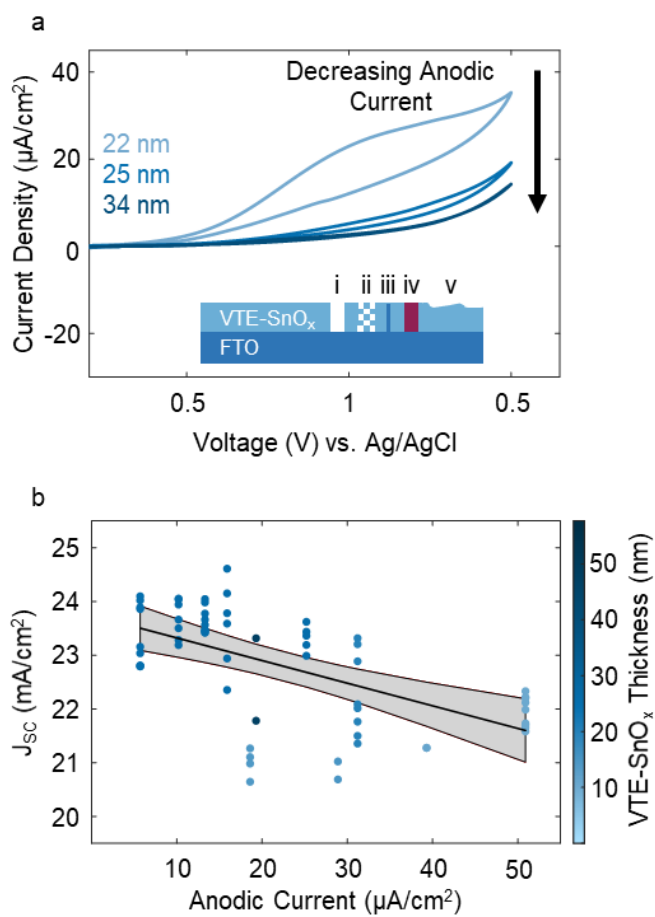

Figure 2. a) Cyclic voltammograms of VTE-SnO$_x$ films of the indicated thicknesses annealed at 195°C. Sources of current leakage include: i-ii) pinholes, iii) conductive filaments, iv) defects, and v) other inhomogeneities shown in the inset schematic. b) Short-circuit current density of final devices plotted against the anodic current at 1.5V vs. Ag/AgCl of a representative film from the same deposition and annealing batch. The shaded grey region is the 95% fit confidence interval. The devices are the same as those of Figure 1b.

Finally, we adapted our electrochemical setup to probe for Sn$^{II}$ defects by switching to a weakly acidic, aqueous methylammonium halide electrolyte (0.1 M). Under these conditions (pH

~ 5.8), we expect $Sn^{II}$ and $Sn^{IV}$ to reduce to $Sn^0$ near -0.9 V and -1.2 V versus Ag/AgCl, respectively.[50] Figure 3a shows photographs of as-deposited and 200°C annealed VTE-SnO$_x$ thin films after 5 minute potentiostatic biasing at the indicated potentials. Significant darkening of the as-deposited sample was observed at potentials ≤ -1.0 V, while darkening did not occur until < -1.2 V for VTE-SnO$_x$ annealed at 200°C. XPS analysis indicates this darkening correlates well with the conversion of VTE-SnO$_x$ to more reduced tin species, including $Sn^0$ (Supplementary Figure S9). The corresponding chronoamperometry data are shown in Supplementary Figure S10. We attribute reduction in the range of -1.0 to -1.2 V vs. Ag/AgCl to the reduction of $Sn^{II}$ species like SnO to $Sn^0$ and at potentials from -1.2 to -1.3 V to reduction of $Sn^{IV}$ in $SnO_2$, in accordance with previous studies at similar pH.[18,51] The presence of tin species with intermediate coordination are suggested by the broadening of the Sn $3d_{5/2}$ peak after electrochemical degradation (Supplementary Figure S9). The effect of methylammonium acidity was verified by conducting control experiments under the same experimental conditions but using a 0.1 M potassium halide electrolyte. As expected, the thresholds for reduction in the VTE-SnO$_x$ films were shifted to more cathodic applied potentials, evidenced by the lack of darkening in the VTE-SnO$_x$ films (Figure S11).[51] The acidic attack enabled by methylammonium agrees with recent studies on the perovskite/selective layer interface.[11,12,52,53] Thus the combined XPS and electrochemical characterization strongly suggests the as-deposited VTE-SnO$_x$ contains significant concentrations of $Sn^{II}$ relative to annealed VTE-SnO$_x$, providing chemical insight into the process and performance variations. Further, this highlights the ability of our electrochemical measurement in methylammonium to reveal the presence of high levels of reactive $Sn^{II}$ by a rapid and simple test.

To augment the electrochemical identification of $Sn^{II}$ defects, we performed XPS to directly measure the surface chemistry of the VTE-SnO$_x$ layers as a function of annealing temperature and confirm the origin of the electrochemical characteristics (e.g. the cathodic onset potential). Figures 3b and 3c show the Sn $3d_{5/2}$-area-scaled Sn $3d_{5/2}$ and VBM spectra of the as-deposited samples compared to samples annealed at the indicated temperatures (see Supplementary Figure S12 for additional Sn$3d_{5/2}$-area-scaled survey and core levels as well as Figures S13-S14 for plots of the raw data). To determine the ratio of Sn:O we used three main techniques. First, we analyzed the Sn $3d_{5/2}$ core level to attempt to discern $Sn^{II}$ from $Sn^{IV}$. However, the close proximity of the SnO and $SnO_2$ Sn $3d_{5/2}$ core level features resulted in a single peak for all samples that we could not confidently decompose,[50,54–57] limiting us to correlate gradual shifts to higher binding energy (Figure 3d) with increased annealing temperatures as a likely indicator for higher ratios of $SnO_2$ to SnO.[58] Second, we analyzed the VBM, which qualitatively shows different valence "fingerprint" features for SnO and $SnO_2$ films.[56–58] Although quantitative interpretation is not possible, the increase in signal in the lower binding energy region of the VBM spectra again suggests that the as-deposited VTE-SnO$_x$ has more $Sn^{II}$ (SnO) character than the annealed films. Note that due to signal to noise issues with the original data, the spectra presented here are from an analogue set of SnO$_x$ samples prepared under the same conditions (see Figure S15 for XPS VBM scans of the SnO$_x$ films corresponding to the core level XPS). Finally, we calculated the O:Sn ratio shown in Figure 3d using the areas of the Sn $3d_{5/2}$ and O 1s peaks scaled by their relative sensitivity factors with contributions from oxygen-bearing C, Ca, Na, and Si contaminants subtracted – assuming they come from the glass substrate due to dust from sample cleaving during preparation (see Supplementary Figures S12-16 and Note III for calculation details). We note that this calculation requires a number of assumptions and therefore has multiple potential sources of error. Nonetheless, these results again suggest that the as-deposited film has a larger ratio of SnO to $SnO_2$ than the other films. Thus, while quantification uncertainties exist, the XPS results all qualitatively indicate a main difference in the VTE-$SnO_x$ is the presence of $Sn^{II}$ defects that are oxidized upon annealing in atmosphere to $Sn^{IV}$, at least at the surface, in good agreement with our electrochemical results (Figure 3a). The challenge of XPS quantification, particularly with elements and chemical states where decomposition is difficult (e.g. Sn $3d_{5/2}$), highlights the ability of electrochemical probes to provide actionable feedback with greater simplicity of measurement and analysis. In the end, both XPS and the chronoamperometry reactivity test reveal that the elimination of $Sn^{II}$ defects play an important role in energetics, which can affect $V_{OC}$, and durability, providing mechanistic insights into the changes we observe in cyclic voltammetry when changing the anneal temperature.

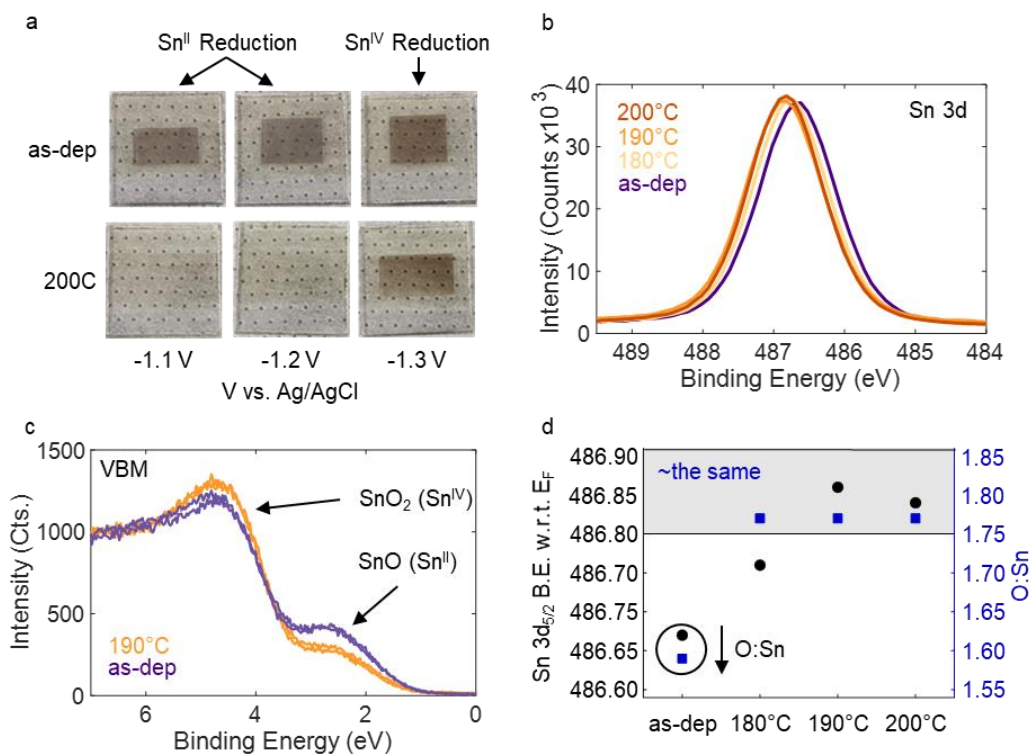

Figure 3. a) Photographs of as-dep and 200°C annealed VTE-$SnO_x$ biased at the indicated potentials (Ag/AgCl reference) in an aqueous 0.1 M methylammonium halide electrolyte visually indicating $Sn^0$ reduction originating from $Sn^{II}$ (-1.1 to -1.2 V) and $Sn^{IV}$ (-1.3 V), inferred from XPS (Figure S9). b) Sn $3d_{5/2}$ and c) X-ray valence band maximum scans (see Figure S15 for VBM scans corresponding to Figure 3b,d). d) Sn $3d_{5/2}$ core level shift and calculated O:Sn ratio (see Supplementary Note III) as a function of annealing temperature.

The electrochemical probes used here – anodic and cathodic CV and potentiostatic stability tests in various electrolytes – and their empirical relationships to device performance enable an informed decision, within minutes, whether to proceed to the next process step or if the process conditions should be changed. Consequently, we now employ these quality control checks in our lab for VTE-SnO$_x$. Coupled with the ability of electrochemical techniques to probe large areas (Figure S17-18), these techniques may contribute to the scaling of MHP fabrication.

In summary, using VTE-SnO$_x$ as a case study, we have revisited the application of simple electrochemical tools to assess the physical and optoelectronic properties of carrier selective contact layers that are relevant to high-performing solid-state optoelectronic devices. Moreover, we have directly correlated various electrochemical properties to final perovskite solar cell performance parameters. Given the critical performance requirements for carrier extraction, efficiency, and interfacial chemical stability in MHP solar cells, we have shown electrochemical tools are particularly well suited to provide rapid and actionable feedback on films incorporated into these devices. It should be noted that while demonstrated on VTE-SnO$_x$, these electrochemical probes are highly adaptable, including for polymeric and hole selective contact materials,[59] and are capable of addressing contact issues at scale. Therefore, they can be used both in the development of new transport layers, by aiding in the troubleshooting process, and/or as a screen for process reproducibility, by permitting comparisons between batches of substrates for perovskite device fabrication. Altogether, the simple electrochemical techniques discussed herein provide a platform for an improved learning cycle, enabling accelerated screening of novel contact layers for specific perovskite compositions to enable well optimized, stable, and high performing devices.

**Supporting Information Available:** Experimental Methods; Supplementary Notes: Electrochemical Apparatus and Design, Calculating VTE-SnO$_x$ Conduction Band Position, XPS-based O:Sn Calculations; Supplementary Figures: Schematic of Apparatus, Defining Cathodic Onset in CV Scans, Comparing Flat-band Voltage to CV Onset, Capacitance-voltage Analysis, Carrier Density, Calculating Conduction Band and Degeneracy, Pinhole Analysis via Chronoamperometry, Anodic Current vs. Shunt Resistance Correlation Plot, Core Level XPS to Show Degradation, Chronoamperometry Data from Electrochemical Degradation, Optical Images of Electrochemically Degraded Films, Sn 3d$_{5/2}$ Scaled XPS of VTE-SnO$_x$ Temperature Series, Raw XPS of VTE-SnO$_x$ Temperature Series, Valence Band XPS, XPS C 1s Deconvolution, Large Area CV and Chronoamperometry of VTE-SnO$_x$; Tables: JV and CV parameters for Figures 1c and 2b.

**Notes**: The authors declare no conflicts of interest.

**Acknowledgements**

This work was supported by the California Energy Commission EPIC program, EPC-16-050 and EPC-19-004. The authors acknowledge the use of facilities and instrumentation at the UC Irvine Materials Research Institute (IMRI), which is supported in part by the National Science Foundation through the UC Irvine Materials Research Science and Engineering Center (DMR-2011967). XPS was performed in part using instrumentation funded in part by the National Science Foundation Major Research Instrumentation Program under grant no. CHE-1338173. Special thanks is afforded to Dr. Ich Tran for assistance in collecting this data. XPS was also performed at the San Diego Nanotechnology Infrastructure (SDNI) of University of California San Diego, a member of the National Nanotechnology Coordinated Infrastructure (NNCI), which is supported by the National Science Foundation (Grant ECCS-1542148). This work was authored in part by the National Renewable Energy Laboratory, operated by Alliance for Sustainable Energy for the US Department of Energy (DOE) under contract no. DE-AC36-08GO28308. Funding for work at